\documentclass{article}
\usepackage{LaThuileFPSpro,cite,amsmath}
\usepackage{hyperref}
\def\MSbar{\hbox{\tiny ${\overline{\rm MS}}$}}
\catcode`\@=11 % This allows us to modify PLAIN macros.
\def\lsim{\mathrel{\mathpalette\@versim<}}
\def\gsim{\mathrel{\mathpalette\@versim>}}
\def\@versim#1#2{\vcenter{\offinterlineskip
        \ialign{$\m@th#1\hfil##\hfil$\crcr#2\crcr\sim\crcr } }}
\catcode`\@=12 % at signs are no longer letters
\begin{document}
%
%\begin{flushright}
%Edinburgh 2008/20
%\end{flushright}

\title{ON THE DETERMINATION OF $|V_{ub}|$ FROM 
INCLUSIVE SEMILEPTONIC B DECAYS\vspace*{15pt}}

\author{
  Einan Gardi      \\ 
\\
  {\em School of Physics, The University of Edinburgh,}\\
{\em Edinburgh EH9 3JZ, Scotland, UK} 
  }
\maketitle

\baselineskip=11.6pt

\begin{abstract}
Precision tests of the CKM mechanism and searches for new physics in the flavour sector require dedicated QCD calculations of decay widths and spectra. 
Significant progress has been achieved in recent years in computing inclusive B decay spectra into light energetic partons.
I briefly review different theoretical approaches to this problem focusing on the determination of~$|V_{ub}|$ from inclusive semileptonic decays and show that this determination is \hbox{robust}. 
The largest uncertainty is associated with the value of the b quark mass.
Finally I present new numerical results in the DGE resummation--based approach, now including ${\cal O}(\beta_0\alpha_s^2)$ corrections.  
The results are presented for all relevant experimental cuts, from which a preliminary average is derived  $|V_{ub}|=(4.30\pm 0.16 ({\rm exp}) ^{+0.09}_{-0.13}({\rm th}) _{- 0.34}^{ +0.39} (m_b) )\cdot 10^{-3}$, where the PDG value of the b quark mass, $m_b^{\MSbar}=4.20\pm0.07$ GeV, is assumed.
\end{abstract}
\newpage
\section{Introduction}

Low--energy precision measurements, in particular precision determination of the CKM parameters and the branching fractions of rare decays, provide many valuable tests of the Standard Model. The resulting constraints on new physics are highly complementary to the direct searches at hadron colliders, and are expected to continue being so throughout the LHC era~\cite{Isidori:2008qp}. 

The experimental effort over the past few years by the B factories and the Tevatron has established the fact that CKM is the main mechanism of flavour and CP violation in the quark sector. This is now a field of precision physics. Measuring deviations from the Standard Model and further strengthening the constraints on new physics will require a continuous experimental effort~\cite{Hashimoto:2004sm,Bona:2007qt,Browder:2007gg,Muheim:2007jk}  
alongside corresponding progress on the theory side.
 
The obvious example of the progress that was made is the precise measurement of the small angle $\beta$ of the Unitarity Triangle, and 
its comparison with the short side of the triangle, $|V_{ub}/V_{cb}|$. 
While the former is directly sensitive to potential CP--violation beyond the Standard Model and can be measured experimentally with high precision without any theoretical input, the latter, being determined by tree--level Weak (semileptonic) decays, is insensitive to new physics, however it heavily relies on theoretical calculations in QCD. 
Currently the measured $\sin(2\beta)$ is not entirely consistent with $|V_{ub}/V_{cb}|$, introducing some tension into global fits~\cite{Charles:2004jd,Bona:2005eu}. This comparison is a crucial element in the big picture.

Experimentally, both $|V_{ub}|$ and $|V_{cb}|$ can be measured either using an exclusive hadronic final state, e.g. 
$\bar{B}\to D^*l \bar{\nu}$  and $\bar{B}\to \pi l \bar{\nu}$ or
by considering the inclusive rate, summing over all hadronic final states subject to some kinematical constraints. The two approaches involve different experimental and theoretical tools and are therefore complementary. 

The lower rate of the $b\to u$ transition 
($|V_{ub}/V_{cb}|^2\sim 1/50$) makes
the $|V_{ub}|$ measurement more challenging. Further difficulties, common to all heavy--to--light decays, lie on the theory side.
The \emph{exclusive} determination of $|V_{ub}|$ requires a theoretical calculation of the form factor using non-perturbative methods 
such as Lattice QCD~\cite{Dalgic:2006dt,Okamoto:2004xg} 
or QCD sum-rules~\cite{Ball:2004ye}, which both have systematic uncertainties that are hard to quantify. 
The Lattice calculations are expected to improve in the future.
In contrast the \emph{inclusive} determination relies primarily on the 
 Heavy Quark Expansion (HQE) and QCD perturbation theory, where a systematic improvement can be achieved and uncertainties are easier to quantify. However, the case of inclusive $b\to u$ decay is further complicated by the fact that the final state is characterized by jet-like kinematics.
 This, in conjunction with the experimental requirement to perform the measurement subject to stringent kinematic cuts (in order to suppress the charm background) implies that to extract $|V_{ub}|$ one needs a precise calculation of the spectrum, not just the total width. 
This provides one of the biggest challenges in Heavy Flavour physics, a subject an intense theoretical effort over the past few years. 
This is the subject of the present talk.

\section{A brief look at inclusive semileptonic $b\to c$ decay}

In order to appreciate the difficulty in determining $|V_{ub}|$, it is useful to compare it to the much favored case of $|V_{cb}|$. Let us therefore briefly review the situation in inclusive semileptonic $b\to c$ decays.

Owing to the high rate of these decays and their distinct characteristics, the B factories provide precise measurements of the  branching fraction, as well as the first few moments, \emph{over the entire phase space}. These truly--inclusive observables can be readily computed using the HQE, for example,
\begin{equation}
\Gamma(\bar{B}\to X_c l \bar{\nu})=
\underbrace{\Gamma(b\to X_c l \bar{\nu};{\mu})}_{\hspace*{-20pt}{\rm{on-shell\, b-quark\,decay\,with\,IR\,cutoff}\hspace*{-20pt}}}
\,+\,
\frac{C_1 \mu_{\pi}^2({\mu})
+C_2 \mu_G^2({\mu})}{m_b^2}
\,\,+\,\,\frac{\left(...\right)}{m_b^3}\,,
\label{HQE}
\end{equation}
where the first term stands for the partonic on-shell $b$-quark decay width $\Gamma(b\to X_c l \bar{\nu})$, computed with an infrared cutoff $\mu$, and other terms, suppressed by powers of the b--quark mass correspond to matrix elements of local operators, computed with an ultraviolet cutoff $\mu$; as usual, the $\mu$ dependence cancels out order by order. Importantly, non-perturbative corrections first appear at order $1/m_b^2$, where there are two non-perturbative matrix elements, the kinetic energy $\mu_{\pi}^2$ and the chromomagnetic energy $\mu_G^2$.

Note that the partonic decay width computed to next-to-leading order (order $\alpha_s$), without a cutoff, already yields a viable approximation to the measured width. This approximation is systematically improved by including non-perturbative corrections as well as higher--order perturbative corrections. There is a continuous progress on both these fronts. Recent fits to the measured moments~\cite{Buchmuller:2005zv,BF_update,Heidelberg} are based on ${\cal O}(\beta_0 \alpha_s^2)$ accuracy \cite{Aquila:2005hq} with power corrections through 
${\cal O}(1/m_b^3)$, computed with leading--order coefficient functions. These fits (performed in the ``kinetic'' or 1S mass schemes) provide a determination of $|V_{cb}|$, $m_b$ and $m_c$ at the $1-2\%$ level together with a determination of $\mu_{\pi}^2$ at the $10\%$ level.

Very recently, complete ${\cal O}(\alpha_s^2)$ corrections have become available in a fully differential form~\cite{Melnikov:2008qs,Pak:2008qt}; ${\cal O}(1/m_b^4)$ corrections have been computed for the first time~\cite{Dassinger:2006md} and also the 
${\cal O}(\alpha_s)$ correction to the coefficient function of 
$\mu_{\pi}^2$ \cite{Becher:2007tk} was calculated. These advances will further push the accuracy, exploiting the potential of the B factory measurements for inclusive $b\to c$ decays. 

\section{The challenge: computing the charmless semileptonic decay spectrum}

Experimental measurements 
~\cite{Bornheim:2002du,Kakuno:2003fk,Limosani:2005pi,Aubert:2005mg,Aubert:2005im,Bizjak:2005hn,Aubert:2007rb}
 of inclusive $b\to u$ decays involve kinematic cuts in order to remove the charm background. Therefore, extracting $|V_{ub}|$ from the data requires theoretical predictions for the fully (triple) differential $\bar{B}\to X_u l \bar \nu$ spectrum.

The main difficulty in computing the spectra of heavy-to-light decays,  $\bar{B}\to X_u l \bar \nu$ or $\bar B\to X_s \gamma$, is in the fact that most events have a jet-like final state where the hadronic system $X$ has a mass which is much smaller than its energy (approximately half of the energy released in the decay, $m_b$). 
The jet kinematics is most easily described in terms of light-cone momenta $P^\pm=E_X\mp p_X$, where a typical event has $P^-$ of order $m_b$ and $P^+$ not far above the QCD scale $\Lambda$.
The decay process involves dynamics on scales that are far apart $P^-\gg P^+$, complicating the perturbative description as well as the separation and parametrization of non-perturbative effects.  

It has long been recognized~\cite{Bigi:1993ex,Neubert:1993um} that an attempt to compute the spectrum in the small--$P^+$ limit by means of the  HQE would run into serious difficulties: the dynamics is dominated by gluons with momenta of ${\cal O}(P^+)$, turning the $\Lambda/m_b$ expansion into a $\Lambda/P^+$ one!
The physical picture behind this breakdown of the expansion is clear: the small lightcone component of the jet is influenced by soft gluon radiation as well as small fluctuations in the momentum carried by the decaying heavy quark. To recover a useful heavy-quark expansion, the dominant effects, those controlled by the scale $P^+$, must be resummed to all orders. This sum gives rise to the well-known ``shape function'', which can be interpreted as the momentum distribution function of the b quark in the B meson. Similar functions appear at higher orders in the heavy--quark expansion.

Recall that in the $b\to c$ case one could make use of the HQE:
the relevant observables were well approximated by perturbation theory and the non-perturbative corrections were restricted to a few \emph{local matrix elements}. 
In contrast, when considering the $b\to u$ case one is required to compute the spectrum at small $P^+$, which is proportional, \emph{already at leading power}, to a non-perturbative object, the ``shape function''. 
This function is defined by the \emph{non-local matrix element}:
\begin{equation}
S(k^+;\mu)=\int_{-\infty}^{\infty}
\frac{dy^-}{4\pi}{\rm e}^{-ik^+y^-}
\left<B\right\vert
{\bar h(y)[y,0]\gamma_+ h(0)}\left|B\right>\,,
\label{SF_def}
\end{equation}
where $k^+$ is a lightcone momentum component ($h$ are heavy--quark effective theory fields, and $[y,0]$ represents a gauge link). This function describes the distribution of momentum carried by the b quark. 
Thus, we observe that instead of having a few unknown non-perturbative matrix elements which enter as power--suppressed corrections, one faces here an unknown \emph{function} already at the leading order in $\Lambda/m_b$!

Described in these terms the problem of computing the spectrum and extracting $|V_{ub}|$ from data may appear hopeless, or at least require a full-fledged non-perturbative approach.
In fact, as we shall see below, the actual situation is significantly better. The partial branching fractions corresponding to experimentally relevant cuts, 
which vary between 20 to 60 percent of the total, can be still estimated reliably with very little non-perturbative input. Moreover, at present, the largest uncertainty in extracting $|V_{ub}|$ is associated with the parametric dependence on the b-quark mass; other uncertainties (e.g. power corrections, Weak Annihilation) can be reduced by further exploiting the data and thus the prospects for an even more precise $|V_{ub}|$ from inclusive decays are high.

In the following I will briefly describe different theoretical approaches that have been developed in the past few years to compute the fully--differential spectrum and thus extract $|V_{ub}|$ from the B factories data. I will not enter into any technical details, just try to give the flavor of the physics involved and the principal differences between the approaches. I will also not cover all the interesting theoretical developments in this area, notably the method to express the $b\to u$ branching fractions directly in terms of the measured photon--energy spectrum in $\bar{B}\to X_s \gamma$, which had some resurrection recently~\cite{Lange:2005qn,Lange:2005xz,Lee:2008vs}, incorporating subleading effects in $\Lambda/m_b$. 

\section{HQE--based structure--function parametrization approach}

The central idea of the HQE--based structure--function parametrization approach, 
which has been recently put forward and implemented by Gambino {\it et. al.}~\cite{Gambino:2007rp}, is to first use the HQE to compute 
carefully--selected observables --- the first few moments of the structure functions --- where this expansion is expected to be most reliable, and then use these observables to constrain the parametrization of the spectrum. 
In this way one bypasses the need to deal with the difficult kinematic region 
$P^-\gg P^+$ where neither the HQE nor perturbation theory converge well.

To explain briefly how the calculation in this approach is set up, let us recall that the triple-differential rate can be written in terms on three hadronic structure functions $W_i(q_0,q^2)$:
\begin{equation}
\frac{d\Gamma}{dq^2dq_0dE_l}=\frac{G_F^2|V_{ub}|^2}{8\pi^3}\left\{
q^2 W_1-\left[2E_l^2-2q_0E_l+\frac{q^2}{2}\right]W_2+q^2(2E_l-q_0)W_3\right\}
\end{equation}
where $q_0$ and $q^2$ are the total leptonic energy and squared invariant mass, respectively.  Ref.~\cite{Gambino:2007rp} computes the shape of the physical structure functions $W_i(q_0,q^2)$ as a convolution \emph{at fixed $q^2$} between non-perturbative distribution functions $F_i(k_+,q^2;\mu)$ and the perturbative (presently the Born-level) structure functions $W^{\rm pert}_i(q_0,q^2)$:
\begin{equation}
W_i(q_0,q^2)=\int dk_+ F_i(k_+,q^2;\mu) W_i^{\rm pert}\left(q_0-\frac{k_+}{2}\left(1-\frac{q^2}{m_bM_B}\right),q^2;\mu\right),
\end{equation}
where the functions $F_i(k_+,q^2;\mu)$ are parametrized and constrained by the first few $q_0$-moments of $W_i(q_0,q^2)$. 

The moments used to set these constraints are computed using the HQE, where the perturbative part currently includes corrections up to ${\cal O}(\alpha_s^2\beta_0)$ \cite{Gambino:2006wk} and power corrections are included through 
${\cal O}(1/m_b^3)$. The separation between the perturbative component and the power--correction terms is based on a hard momentum cutoff 
($\mu=1$ GeV) in the ``kinetic scheme'', which has the advantage that the input parameters (in particular $m_b$ and $\mu_{\pi}^2$) can be taken directly from fits to the $b\to c$ moments.  

Similarly to the $b\to c$ analysis both the power expansion and the perturbative expansion\footnote{At present fixed--order ${\cal O}(\alpha_s^2\beta_0)$ expressions are used. It is fair to say that the generalization of the hard cutoff approach beyond the level of a single gluon (possibly dressed) is difficult to implement.} can be improved once higher--order corrections are known. The use of a hard cutoff on the gluon energy (the ``kinetic scheme'') eliminates the sensitivity to multiple soft emission rendering the expansion better convergent. Nevertheless a single--logarithmic collinear divergence persists, and can in principle be resummed. 

In this approach the parameters in $F_i$, $i=1,2,3$ are fixed a new
at each given value of $q^2$, based on the moment constraints. 
Thus, the way these parameters vary with $q^2$ is indirectly determined by the HQE.  
This issue becomes crucial at large $q^2$, where the HQE breaks down: the final--state hadronic system is then soft. 
While the contribution from this phase--space region is small (it is power suppressed) the spectrum there is clearly not well under control. Ref.~\cite{Gambino:2007rp} provides an interesting analysis of the breakdown of the HQE in this region and relates it to the presence of ${\cal O}(1/m_b^3)$ Weak Annihilation contributions, centered at $q^2\sim m_b^2$. The contributions from the large--$q^2$ region are parametrized making a conservative estimate of their size. 
Even then the impact of the Weak Annihilation contributions on the average value of $|V_{ub}|$ is just $\sim 3\%$~\cite{HFAG_PDG08}, which is less than the parametric uncertainty due to $m_b$ dependence.
Having said that, Ref.~\cite{Gambino:2007rp} has clearly demonstrated 
that further experimental input on the $q^2$ distribution and moments, measured separately for charged and neutral B mesons, would be important for reducing the uncertainty on $|V_{ub}|$. 

To summarize, the approach of Ref.~\cite{Gambino:2007rp} is cautious: it uses the well--understood (and well tested!) theoretical framework of the HQE for carefully selected moments, and 
assumes very little beyond that. It relies however, on extensive parametrization, dealing with three non-perturbative functions $F_i$ of \emph{two kinematic variables}, whose properties are unknown. The authors of Ref.~\cite{Gambino:2007rp} therefore took special care to consider a large class of functions and further devised means to assess whether this class is large enough. In this way they managed to provide a reliable prediction for the triple differential spectrum over the entire phase space without dealing directly with the difficult kinematic region where
the hadronic system is jet-like.
In the following I present other theoretical approaches that instead consider directly this region.  

\section{Shape--function approach\label{sec:SF}}

The shape--function approach by Neubert and collaborators~\cite{Bosch:2004th,Lange:2005yw} deals directly with the important kinematic region where $P^+\ll P^-$. This is done by establishing a modified expansion in inverse powers of the mass, where at each order the dynamical effects that are associated with soft gluons, $k^+\sim {\cal O}(P^+) \sim {\cal O} (\Lambda)$, 
are summed into non-perturbative shape functions. At leading power there is one such function, the momentum distribution function defined in (\ref{SF_def}) above; beyond this order there are several different functions with additional fields insertions. 
To extend the calculation beyond this particular region, it is constructed to match the standard HQE when integrated over a significant part of the phase space. In this way two systematic expansions in inverse powers of the mass are used together.

The modified expansion in shape functions is developed, following the Soft Collinear Effective Theory (SCET) methodology~\cite{Bauer:2001yt,Bauer:2002uv,Beneke:2002ph}, for the 
 particular kinematic region where the final--state is jet-like $P^-\simeq m_b$, $P^+\simeq \Lambda$, and thus $m_X=\sqrt{m_b\Lambda}$, the region into which a large fraction of the events fall.  
The large (parametric) hierarchy between these scales implies loss of quantum--mechanical coherence between the respective excitations, leading to factorization into three different subprocesses ~\cite{Korchemsky:1994jb} (see also \cite{Bauer:2001yt,Bosch:2004th,Gardi:2004ia}). The result, at leading power in $\Lambda/m_b$, can be expressed as a convolution integral~\cite{Lange:2005yw}:
\begin{equation}
\label{conv}
\frac{d\Gamma}{dP^-dP^+dE_l}\simeq
H(y,\mu) \int_0^{P^+} d k^+ ym_b J\left(ym_b (P^+-k^+),\mu\right)\, S(k^+,\mu)
\end{equation}
where $y\equiv (P^--P^+)/(M_B-P^+)$. Here $H$ stand for \emph{hard}, depending on momenta of ${\cal O}(m_b)$, $J$ for \emph{jet}, depending on momenta of ${\cal O}(\sqrt{m_b\Lambda})$ and $S$ for \emph{soft}, depending on momenta of ${\cal O}(\Lambda)$ and is therefore considered as a non-perturbative object, to be parametrized. 
Similar factorization formulae apply at subleading powers in 
$\Lambda/m_b$, leading to the following expression for the differential width:
\begin{equation}
\label{eq:SF_subleading}\frac{d\Gamma}{dP^-dP^+dE_l}=H J \otimes {S} + \frac{\sum H_n J_n\otimes {S_n}}{m_b} +\cdots
\end{equation}
The matching into the standard HQE translates into constraints on the moments of the shape functions. 

The authors of Ref.~\cite{Lange:2005yw} have defined the separation between the leading term and the power corrections using a factorization procedure that is based on dimensional regularization. They parametrize the shape functions directly\footnote{Note that in this way the (known) evolution properties of the shape--function are not being used.} at the intermediate (jet) scale 
$\mu\sim \sqrt{m_b \Lambda}$ (in practice $\mu=1.5$ GeV is used) and thus avoid ever dealing with softer momentum scales.

Both the hard and the jet functions are computed in perturbation theory. 
Owing to the presence of well-separated scales, the jet energy 
${\cal O}(m_b)$ and its mass ${\cal O}(\sqrt{m_b\Lambda})$ the perturbative expansion contains large Sudakov logarithms. Ref.~\cite{Lange:2005yw} resum these logarithms to all orders with high logarithmic accuracy (next--to--next--to--leading log, NNLL). 
The hard coefficient function is currently computed at 
${\cal O}(\alpha_s)$. 

Variation of the matching scale is used to estimate missing higher--order corrections. This translates into about $3-4\%$ uncertainty on the average value of $|V_{ub}|$. 

The functional forms of the shape functions are unknown, and they are  therefore parametrized. The first two moments of the leading shape function are reasonably well constrained: they are fixed by $m_b$ and $\mu_{\pi}^2$ respectively; higher moments are not well constrained. Subleading shape functions are difficult to constrain.

To summarize, the method of Ref.~\cite{Lange:2005yw} makes extensive use of the available theoretical tools, employing consistently two expansions that are valid in two different kinematic regimes: the expansion in shape functions, valid for the typical final--state momentum configuration, and the standard HQE, valid for the fully integrated width. Further to that, Sudakov resummation for the jet-scale logarithms is employed at NNLL accuracy. By using a relatively high factorization scale the perturbative calculation remains insensitive to the infrared, and converges well. 
All ingredients, the power expansions as well as the perturbative ones, can in principle be improved systematically by including higher--order terms. 

The motivation to go to subleading powers in a completely general way, however, has a price: one needs to parametrize several different subleading shape functions on which there is no theoretical control nor experimental input. Despite this, the authors of Ref.~\cite{Lange:2005yw} have demonstrated that the experimentally--relevant partial branching fractions remain under good control: their sensitivity to the unknown higher moments of the leading shape function as well as to the unknown functional form of the subleading shape functions is small: the estimated effect of these unknowns on the average value of $|V_{ub}|$ is less than $1\%$! The largest uncertainty in the determination of $|V_{ub}|$ is the parametric one, owing to the strong dependence on $m_b$.

A common feature of the two approaches described so far is the
extensive use of parametrization of non-perturbative functions. The fact that these functions all have a clear field--theoretic definition does not presently help in their parametrization. 
To improve on that one needs to avoid  introducing an explicit cutoff $m_b\gg\mu\gg\Lambda$. 
This is indeed possible.
It is well--known that inclusive observables, such as moments of decay spectra, are \emph{infrared--safe} observables. In other words, in the absence a cutoff soft--gluons divergences cancels out in the sum of real and virtual diagrams, making the moments finite at any order in perturbation theory.  
In the following I shall describe a resummation--based approach, where a cutoff is not used and consequently one relies less on parametrization.

\section{Resummation--based approach}

The approach of Refs.~\cite{Gardi:2004ia,Andersen:2005bj,Andersen:2005mj,Andersen:2006hr} 
uses resummed perturbation theory in moment--space to provide a perturbative calculation of the on-shell decay spectrum in the entire phase space without introducing any external momentum cutoff; non-perturbative effects are taken into account as power corrections in moment space. 
Resummation is applied to both the `jet' function and the `soft' (quark distribution) function, dealing directly with the double hierarchy of scales characterizing the decay process. Consequently, the shape of the spectrum in the kinematics region where the final state is jet-like is largely determined by a calculation, and less by parametrization. 

The resummation method employed, DGE\footnote{Dressed Gluon Exponentiation (DGE) is a general resummation formalism for inclusive distributions near a kinematic threshold~\cite{DGE_review}. It goes beyond the standard Sudakov resummation framework by incorporating renormalon resummation in the calculation of the exponent. This has proven effective~\cite{Cacciari:2002xb,Gardi:2001ny} in extending the range of applicability of perturbation theory nearer to threshold and in identifying the relevant non-perturbative corrections in a range of applications~\cite{DGE_review,Gardi:2001di,Cacciari:2002xb,Gardi:2001ny,Gardi:2007ma}.}, combines Sudakov and renormalon resummation. Sudakov logarithms are resummed with high logarithmic accuracy (NNLL) for both\footnote{The `jet' logarithms are similar to those resummed 
in the approach of Ref.~\cite{Lange:2005yw}; there however `soft' logarithms are not resummed.} the `jet' and the `soft' functions.

Renormalon resummation is an essential element in implementing a consistent separation between perturbative and non-perturbative corrections at the power level. 
Refs.~\cite{Andersen:2005bj,Andersen:2005mj,Andersen:2006hr} have adopted the Principal Value procedure to regularized the Sudakov exponent and thus \emph{define} the non-perturbative parameters.
This is in full analogy~\cite{Gardi:1999dq} with the way a momentum cutoff is conventionally used.
Most importantly, this definition applies to the would-be $1/m_b$ ambiguity of the `soft' Sudakov factor, which cancels exactly~\cite{Gardi:2004ia} against the pole--mass renormalon when considering the spectrum in physical hadronic variables. The same regularization used in the Sudakov exponent must be applied in the computation of the b-quark pole mass\footnote{In Eq.~(\ref{inv_Mellin}) below the cancellation of the renormalon ambiguity involves Sudakov factor of Eq.~(\ref{fact}), on the one hand, and $\bar{\Lambda}$ on the other.}.
This vital mechanism is absent in a fixed--logarithmic--accuracy procedure (as employed for example in~\cite{Aglietti:2007ik,Aglietti:2005eq}) leading to an uncontrolled shift of the entire spectrum with $P^+/P^-$.

Aiming to provide a good description of the spectrum in the kinematic region where there is a large hierarchy between the lightcone momentum components, $P^+\ll P^-$, it proves useful to consider the moments with respect to the lightcone--component ratio\footnote{Note that we consider here the moments of the \emph{fully differential width}~\cite{Andersen:2005mj}: the moments remain differential with respect to the large lightcone component $p^-$ as well as the lepton energy $E_l$. This is essential for performing soft gluon resummation. This issue has also been discussed in Ref.~\cite{Aglietti:2005eq}.}:
\begin{equation}
\label{mom}
\frac{d\Gamma_N(p^-,\,E_l)}{dp^-\,dE_l}\,\equiv \,\int_0^{p^-} {dp^+}\,\left(1-\frac{p^+}{p^-}\right)^{N-1} \frac{d\Gamma(p^+,\,p^-,\,E_l)}{dp^+\,dp^-\,dE_l},
\end{equation}
where the partonic lightcone momentum components $p^{\pm}$ are related to the hadronic ones by: $p^{\pm}=P^{\pm}-\bar{\Lambda}$, where $\bar{\Lambda}=M_B-m_b$ is the energy of the light--degrees--of--freedom in the meson. 

Note that in (\ref{mom}) large moment index corresponds to 
the limit of interest, jet kinematics: the main contribution to the integral for $N\to \infty$ comes from the region where $p^+/p^-\to 0$. 
For large $N$ one identifies three characteristic scales, \emph{hard} \,
${\cal O}(p^-)$, \emph{jet}\, ${\cal O}(p^-/\sqrt{N})$ and \emph{soft}\, ${\cal O}(p^-/N)$. 
In this limit, and up to $1/N$ corrections, the moments \emph{factorize}~\cite{Korchemsky:1994jb,Bauer:2001yt,Bosch:2004th,Gardi:2004ia} to all orders as follows\footnote{Note that this factorization formula maps directly  onto eq.~(\ref{conv}) above, where the convolution integral turns into a product in moment space.}:
\begin{eqnarray}
\label{fact}
\frac{d\Gamma_N(p^-,\,E_l)}{dp^-\,dE_l}\,=
H(p^-,\,E_l)\, \underbrace{J\big({p^-}/{\sqrt{N}},\,\mu\big)\, S_{b}\big({p^-}/{N},\,\mu\big)}_{{\rm Sud}(p^-,\, N)}\, + \,{\cal O}(1/N),
\end{eqnarray}
where the factorization--scale ($\mu$) dependence cancels \emph{exactly} in the product in the Sudakov factor ${\rm Sud}(p^-,\, N)$.

\begin{figure}[htb]
\vspace*{5cm}
\begin{center}
\hspace*{40pt}  
\includegraphics{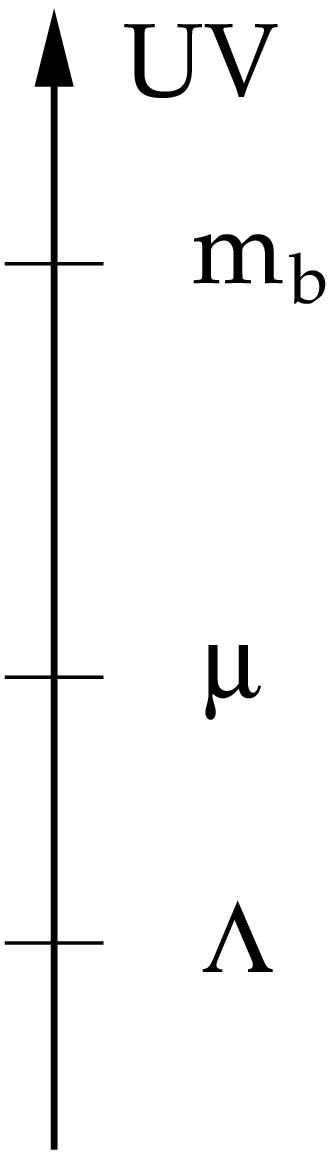}
\hspace*{120pt}
\includegraphics{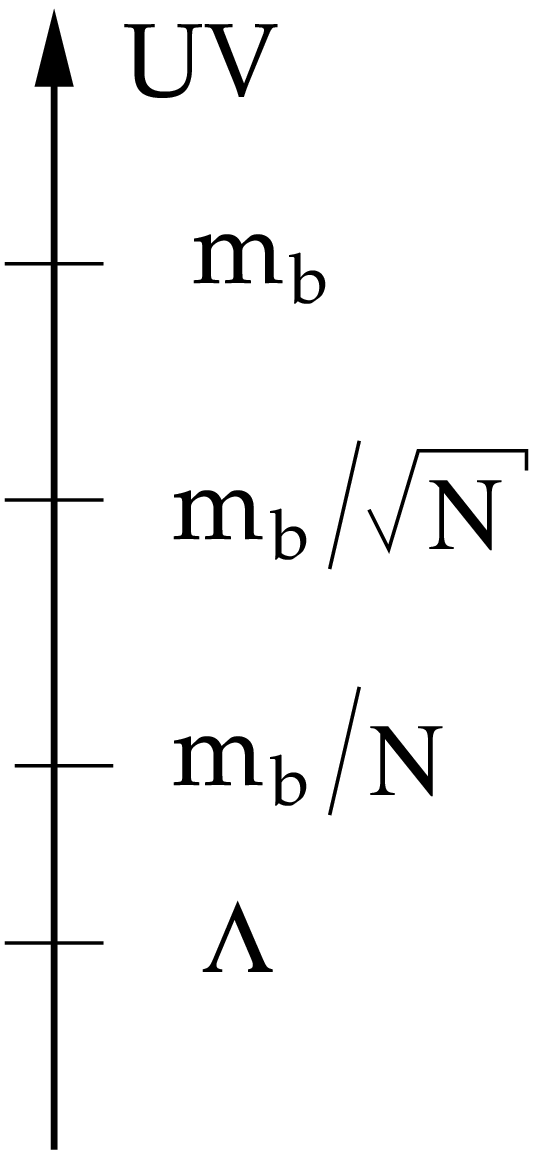}
\end{center}
  \caption{\it
    The hierarchy of scales underlying factorization as conceived in a cutoff--based approach (left) {\it vs.} the moment--space resummation--based approach (right). 
    \label{fig:hierarchy} }
\vspace*{1cm}
\end{figure}

To use perturbation theory one must consider $p^-$  large enough, such that even the `soft' scale $p^-/N$ is sufficiently large compared to the QCD scale $\Lambda$. 
This hierarchy of scales is illustrated in 
figure \ref{fig:hierarchy}. Here one should note a subtle but important distinction from  the shape--function approach discussed above, where it was a priori assumed that the ``soft'' scale (here $p^-/N$) is ${\cal O}(\Lambda)$, prohibiting any perturbative treatment of the corresponding dynamical subprocess. Here instead we wish to compute $S_{b}\big({p^-}/{N},\,\mu\big)$ --- the quark distribution inside \emph{an on-shell heavy quark}~\cite{Korchemsky:1992xv,Gardi:2005yi} --- in perturbation theory, as a basis for the description of the physical distribution, the 
quark distribution in the B meson $S_{B}\big({p^-}/{N},\,\mu\big)$.
Because $S_{b}\big({p^-}/{N},\,\mu\big)$ is infrared safe $S_{B}\big({p^-}/{N},\,\mu\big)$ only differs from $S_{b}\big({p^-}/{N},\,\mu\big)$ by power corrections, powers of $N\Lambda/p^-$. Eventually, at 
$N\gg p^-/\Lambda$ all these powers become relevant, recovering the ``shape function'' scenario. Refs.~\cite{Andersen:2005bj,Andersen:2005mj,Andersen:2006hr}
therefore parametrize these power corrections. 
It should be noted that experimentally--relevant branching fraction are not so sensitive to the high moments, and therefore the effect of these power corrections is small. The resulting effect on $|V_{ub}|$ is at the sub-percent level. 

Factorization facilitates the resummation of Sudakov logarithms, the corrections that dominate the dynamics at large $N$~\cite{Korchemsky:1994jb,Akhoury:1995fp,Aglietti:2005eq,Bauer:2001yt,Bosch:2004th,Gardi:2004ia,Aglietti:2001br}.
There is, however, another class of large corrections which is always important at high orders: these are running--coupling corrections, or renormalons.  In the approach of Refs.~\cite{Andersen:2005bj,Andersen:2005mj,Andersen:2006hr} the Sudakov exponent is computed as a Borel sum, facilitating simultaneous resummation of 
Sudakov logarithms and running--coupling corrections. The Sudakov factor takes the form:
\begin{align}
\begin{split}
\label{Sud_DGE} &{\rm Sud}(p^-,\,N)=\exp\bigg\{\frac{C_F}{\beta_0}
\int_0^{\infty}\frac{du}{u}\,\left(\frac{\Lambda}{p^-}\right)^{2u}\,
\bigg[B_{\cal S}(u)G(2u,N)-B_{\cal J}(u)G(u,N)\bigg]\bigg\}\,, \\
&
{\rm where}\qquad 
G(u,N)=\Gamma(-u)
\left(\frac{\Gamma(N)}{\Gamma(N-u)}-\frac{1}{\Gamma(1-u)}\right)\,.
\end{split}
\end{align}
Here~$B_{\cal S}(u)$ and $B_{\cal J}(u)$ are the Borel representations of the Sudakov anomalous dimensions of the quark distribution and the jet function, respectively. These functions are  known \cite{Gardi:2005yi} to NNLO, ${\cal O}(u^2)$, facilitating Sudakov resummation with next--to--next--to--leading logarithmic accuracy~\cite{Andersen:2005bj}.  

Beyond that, the \emph{analytic structure} of the integrand is indicative of power corrections. The Sudakov exponent has renormalon singularities at integer and half integer values of~$u$, except where $B_{\cal S,\, J}(u)$  vanish. The corresponding ambiguities, whose magnitude is determined by the residues of the poles in~(\ref{Sud_DGE}), are enhanced at large $N$ by powers of $N$. They indicate the presence of non-perturbative power corrections with a similar $N$ dependence. These power corrections \emph{exponentiate} together with the logarithms. 
By evaluating the Borel integral, rather than expanding it, Refs.~\cite{Andersen:2005bj,Andersen:2005mj,Andersen:2006hr} make use of this additional information, defining the perturbative part of the exponent via the Principal Value (PV) prescription, and then parametrizing the dominant power corrections.
 
So far only non-perturbative corrections that are leading in the large $N$ limit, $(N\Lambda/p^-)^k$ for any $k$, have been taken into account in this approach.
These power corrections are the non-perturbative content of the leading shape function in the approach of Sec.~\ref{sec:SF}. 
${\cal O}(1/N)$ effects 
corresponding to subleading shape functions in (\ref{eq:SF_subleading})
are only accounted for in Refs.~\cite{Andersen:2005bj,Andersen:2005mj,Andersen:2006hr} at the perturbative level. In principle, however, subleading non-perturbative effects can also be parametrized as power corrections in moment space. This would be worthwhile doing at the point where the constraints on the leading power terms would be sufficiently tight.

Moment space proves convenient for resummation and parametrization of power corrections, but at the end of the day one needs the spectrum in momentum space.
The fully differential spectrum in hadronic variables is obtained by an inverse Mellin transform ({\it cf.} (\ref{mom})):
\begin{equation}
\label{inv_Mellin}
\frac{d\Gamma(P^+,\,P^-,\,E_l)}{dP^+\,dP^-\,dE_l}
\,=\left. \int_{c-i\infty}^{c+i\infty} \frac{dN}{2\pi \, i}\,\left(1-\frac{p^+}{p^-}\right)^{-N}
\frac{1}{p^-}\,\frac{d\Gamma_N(p^-,\,E_l)}{dp^-\,dE_l}
\right\vert_{p^{\pm}=P^{\pm}-\bar{\Lambda}}
\end{equation}
where the integration contour runs parallel to the imaginary axis, to the right of the singularities of the integrand. 

To summarize, the moment--space resummation approach of Refs.~\cite{Gardi:2004ia,Andersen:2005bj,Andersen:2005mj,Andersen:2006hr} allows to compute the fully--differential spectrum in the entire phase space as an infrared--safe quantity, without introducing any explicit cutoff scale.
Non-perturbative effects are treated as power corrections, where the parameters are defined using the Principal Value prescription.
This approach thus maximizes the predictive power of perturbation theory, and minimizes the role of parametrization. In the next section we shall have a quick look at the resulting phenomenology. In particular, we will present here for the first time numerical results that are based on matching the resummation formula to  ${\cal O}(\beta_0\alpha_s^2)$, incorporating the results of Ref.~\cite{Gambino:2006wk}.

\section{$|V_{ub}|$ by DGE including ${\cal O}(\beta_0\alpha_s^2)$ corrections}

So far the calculation of the partial branching fractions from which $|V_{ub}|$ was extracted, has been based on a NLO result: although the jet and the soft functions were resummed with NNLL accuracy, the hard coefficient function $H(p^-,\,E_l)$ in (\ref{fact}), corresponding to constants and $1/N$ suppressed terms at large $N$, was only known to NLO, 
${\cal O}(\alpha_s)$~\cite{DeFazio:1999sv}. In a recent paper \cite{Gambino:2006wk} we have computed analytically the running--coupling corrections, which are the dominant corrections at the NNLO. Both real and virtual ${\cal O}(\beta_0 \alpha_s^2)$ corrections are now available. 

Very recently I have completed the task of matching the resummed 
triple--differential rate to the new ${\cal O}(\beta_0 \alpha_s^2)$ corrections, and implemented it into the C$++$ DGE program. The details of the matching procedure will be published separately. The new version of the program (Version 2.0) is available at \cite{new_DGE_program}. Preliminary results based on this new version will be presented below.

Prior to describing the new results a comment is due concerning the 
way the partial branching fractions are computed, which has changed between the old and new implementations~\cite{new_DGE_program}.
In the old version the triple differential width, normalized as
$1/\Gamma_0 d\Gamma/dP^+dP^-dE_l$ ($\Gamma_0$ is the Born--level width) was integrated over the relevant phase-space, and then divided by a normalization factor corresponding to a similar integral \emph{over the entire phase space}. In the new version, I apply the same procedure but this time evaluating at each point in phase space the (perturbatively) normalized rate $1/\Gamma_{\rm total} d\Gamma/dP^+dP^-dE_l$ instead of $1/\Gamma_0 d\Gamma/dP^+dP^-dE_l$. This implies that the expression for $\Gamma_0/\Gamma_{\rm total}$ has been expanded, and multiplied into the hard matching coefficient. Finally the new hard matching coefficient is truncated at the required order, $\alpha_s$ (NLO) or $\beta_0 \alpha_s^2$ (NNLO). 
When working at NLO this amounts to an ${\cal O}(\alpha_s^2)$ difference with respect to the previous calculation, which is not small numerically (it is comparable to the $\beta_0\alpha_s^2$ term in the matching coefficient, and has the opposite sign). The new formulation is theoretically favored as it leads to smaller renormalization--scale dependence\footnote{Renormalization scale dependence appears in our formulation only through the hard matching coefficients, as running coupling corrections are resummed in the jet and soft functions.}.

Using the 2007 PDG value for the short--distance b-quark mass,
\begin{equation}
m_b^{\MSbar}=4.20 \pm 0.07 \,\,{\rm GeV}\,
\label{m_b_PDG}
\end{equation}
I have computed the normalized partial widths 
\begin{equation}
R_{cut}=
\frac{\Gamma({\bar B}\to X_u l {\bar \nu}; \,{\rm cut})}{\Gamma_{\rm total}({\bar B}\to X_u l {\bar \nu})}
\end{equation}
for the specific cuts used by HFAG~\cite{HFAG_PDG08} to extract $|V_{ub}|$ based on the measurements in~\cite{Bornheim:2002du,Kakuno:2003fk,Limosani:2005pi,Aubert:2005mg,Aubert:2005im,Bizjak:2005hn,Aubert:2007rb}.
The results are summarized in table \ref{Rcut_tab}. Note that the lepton energy cut is sometimes applied in the $\Upsilon(4S)$ frame rather than the B rest frame, involving a boost of $\beta=0.064$; this is indicated in the table and taken into account in the calculation.
 \begin{table}[t]
  \centering
  \caption{ \it Computed values of $R_{\rm cut}$ for different experimentally relevant cuts. The collumn on the left described the cuts. 
For $R_{\rm cut}^{\rm NLO}$ we present the central value only, while for $R_{\rm cut}^{\rm NNLO}$ the errors are broken into (an asymmetric) theory error (which includes parametric uncertainty in the input value of $\alpha_s$; renormalization scale uncertainty; power corrections associated with the quark distribution function; and Weak Annihilation effect) and parametric uncertainty in the input value of $m_b$ according to (\ref{m_b_PDG}). 
    }
  \vskip 0.1 in
  \begin{tabular}{|c|c|c|c|} \hline
&&&\\
     cut& Ref. & $R_{\rm cut}^{\rm NLO}$ &  $R_{\rm cut}^{\rm NNLO}$ \\
&&&\\
    \hline
    \hline
&&&\\
   $E_l>2.1$ GeV &\cite{Bornheim:2002du} &0.234&$0.223\,^{+0.021} _{-0.013}\,^{+0.024} _{-0.024}(m_b)$\\&&&\\
% run with   
$E_l^{\Upsilon(4S)}>1 $ GeV;
$m_X<1.7$ GeV; $q^2>8$ GeV  &\cite{Kakuno:2003fk}   &0.372&$0.366\,^{+0.021} _{-0.007}\,^{+0.016} _{-0.017}(m_b)$\\&&&\\
  $E_l^{\Upsilon(4S)}>1.9$ GeV &\cite{Limosani:2005pi} &0.389&$0.374\,^{+0.021} _{-0.013}
\,^{+0.023} _{-0.022}(m_b)$\\&&&\\
   $E_l^{\Upsilon(4S)}>2.0$ GeV  &\cite{Aubert:2005mg}   &0.311&$0.301
\,^{+0.020} _{-0.015}\,^{+0.024} _{-0.025}(m_b)$\\&&&\\
  $E_l>2.0$ GeV; $S_h^{\max}<3.5$ GeV$^2$  &\cite{Aubert:2005im}   &0.239&$0.232\,^{+0.020} _{-0.013}\,^{+0.023} _{-0.022}(m_b)$\\&&&\\
   $E_l^{\Upsilon(4S)}>1.0$ GeV; $m_X<1.7$ GeV  &\cite{Bizjak:2005hn}   &0.658&$0.628
\,^{+0.026} _{-0.017}\,
^{+0.055} _{-0.059}(m_b)$\\&&&\\
   $E_l^{\Upsilon(4S)}>1.0$ GeV; $m_X<1.55$ GeV   &\cite{Aubert:2007rb}   &0.559&$0.532
\,^{+0.028} _{-0.023}\,^{+0.069} _{-0.071}(m_b)$\\&&&\\
    \hline
  \end{tabular}
  \label{Rcut_tab}
\end{table}

As shown in table \ref{Rcut_tab} a significant contribution to the uncertainty is due to the parametric dependence on $m_b$ for which we have taken the conservative range of (\ref{m_b_PDG}). 
Other uncertainties we take into account  (all summed up in quadrature
in table \ref{Rcut_tab}) are: 
\begin{itemize}
\item{} 
Parametric uncertainty in the input value of $\alpha_s$, where we take $\alpha_s^{\MSbar}(M_Z)=0.1176\pm 0.020$.
\item{} Power corrections associated with the quark distribution function, estimated by varying the $u=3/2$ renormalon residue as well as the power terms based on the parametrization presented in Sec. 4.3 in~\cite{Andersen:2006hr}. We take $(C_{3/2},f^{\rm PV})=(1,0)$ as default and determine the uncertainty by considering the case $(C_{3/2},f^{\rm PV})=(6.2,0.3)$.
\item{} Weak Annihilation effect. We assume that Weak Annihilation effects can increase the width by up to 2\%. This error is taken as unidirectional.
\item{} The residual dependence on $\mu$ in the matching coefficient is used to estimate higher--order perturbative corrections. We vary it from $\mu=m_b/2$ to $\mu=2m_b$, where the central value is taken at $\mu=m_b$.
\end{itemize}
Note that the NNLO result for each of the cuts is consistent within errors with the NLO one. For NLO we only quote the central values; the errors are similar to those at NNLO. In particular, considering here the normalized $R_{\rm cut}$ computed by integrating  
$1/\Gamma_{\rm total} d\Gamma/dP^+dP^-dE_l$
 the renormalization--scale dependence is low (1--2\% on $|V_{ub}|$) already at NLO and there is no significant improvement going to NNLO. 

\begin{figure}[htb]
\vspace*{8cm}\hspace*{380pt}
\includegraphics{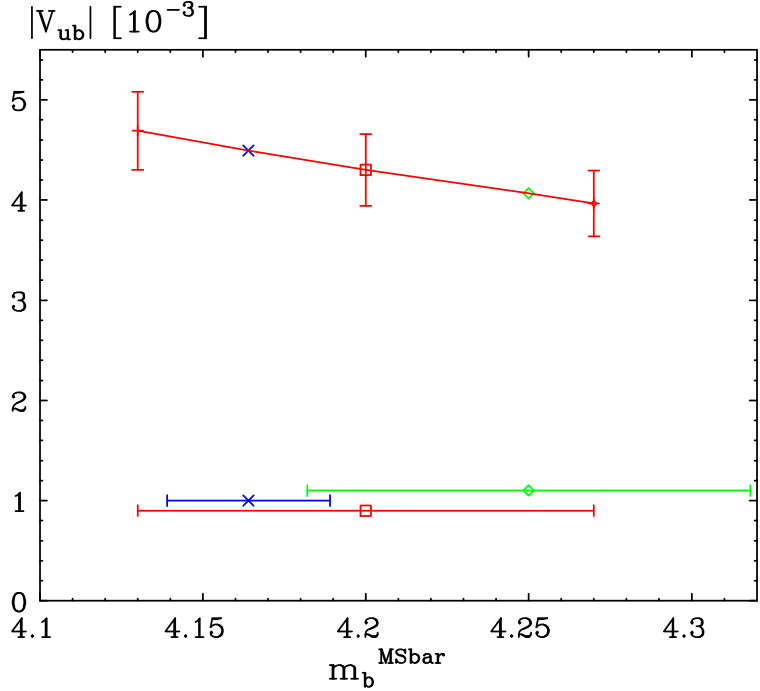}
  \caption{\it
    The extracted value of $|V_{ub}|$, averaged over different measurements with different kinematic cuts, as function of the b quark mass, $m_b^{\MSbar}$. The quark mass serves as an input to the calculation of the partial width; it affects the partial width thought the total width, $\sim m_b^5$, and through $R_{\rm cut}$,  both acting in the same direction. The calculation of $R_{\rm cut}$ is done by DGE including ${\cal O}(\beta_0\alpha_s^2)$ NNLO corrections. 
The vertical error bars are based on the remaining sources of uncertainty, theoretical and experimental, added in quadrature. 
The wide red horizontal bar at the bottom shows the 1-sigma range of the
PDG world average value (\ref{m_b_PDG}) setting the range of $m_b$ values we consider. Just above it we present two specific determinations of the mass: the one extending to the right (green) based on a HQE--based fit to inclusive moments of $b\to c$ decays~\cite{Buchmuller:2005zv,BF_update}, converted to $\overline{\rm MS}$~\cite{Neubert:2008cp}, and the other (blue) based on a recent precise determination~\cite{Kuhn:2007vp} using the total
cross section in $e^+e^-\to{hadrons}$ near the bottom production threshold.    \label{fig:Vub_mass_dependence} }
\vspace*{1cm}
\end{figure}

Next, to extract $|V_{ub}|$ the experimental partial 
branching fractions~\cite{Bornheim:2002du,Kakuno:2003fk,Limosani:2005pi,Aubert:2005mg,Aubert:2005im,Bizjak:2005hn,Aubert:2007rb} can be directly compared to the theoretical calculation:
\begin{equation}
\label{DeltaB}
\Delta{\cal B}({\bar B}\to X_u l {\bar \nu}; \,{\rm cut})=\tau_B \times \Gamma_{\rm total}  (\bar{B}\to X_ul \bar{\nu}) \times R_{\rm cut}.
\end{equation}
Using the calculation of Sec. 2 in \cite{Andersen:2005mj} 
with  PDG value of $m_b$ (\ref{m_b_PDG}) we get the following value for the total width:
\begin{equation}
\label{total}
\frac{1}{|V_{\rm ub}|^2}\, \Gamma_{\rm total} (\bar{B}\to X_ul \bar{\nu})=67.3 \pm 5.4 {\rm ps}^{-1} \,.
\end{equation}
Using the updated world average value of the B-meson life time, $\tau_B=1.573\, {\rm ps}$, together with (\ref{total}) and the $R_{\rm cut}$ values of table \ref{Rcut_tab} we obtain for $|V_{ub}|$ the values quoted in table \ref{Vub_tab}.
\begin{table}[t]
  \centering
  \caption{ \it  Extracted  values of $|V_{ub}|$  based on the measured partial branching fractions (see quoted references), using Eq. (\ref{DeltaB}) with the $R_{\rm cut}$ values of table \ref{Rcut_tab} and the total width of Eq.~(\ref{total}).
 The errors quoted for NNLO are experimental (statistic and systematic raised in quadrature); theoretical, through $R_{cut}$; and parametric dependence on $m_b$ in both the total width and through $R_{cut}$.  
    }
  \vskip 0.1 in
  \begin{tabular}{|c|c|c|c|} \hline
&&&\\
cut
& Ref. 
& \begin{tabular}{l}
$|V_{ub}|$
\\ NLO
\end{tabular}
& 
\begin{tabular}{l}
 $|V_{ub}|$\\ 
NNLO
\end{tabular}\\
&&&\\
    \hline
    \hline
&&&\\
   $E_l>2.1$ GeV  
&\cite{Bornheim:2002du} 
&3.64
&$3.73\pm 0.44 ({\rm exp})  ^{+ 0.11} _{-0.16} ({\rm th}) _{ -0.33}  ^{+0.39} (m_b) $\\&&&\\
  %run with 
$E_l^{\Upsilon(4S)}>1 $ GeV;
% measured with  $E_l^{\Upsilon(4S)}>1.2 $ GeV;
$m_X<1.7$ GeV; $q^2>8$ GeV  
&\cite{Kakuno:2003fk}   
&4.33
&$4.36\pm 0.46 ({\rm exp})  ^{+ 0.04} _{-0.12} ({\rm th}) _{ -0.26}  ^{+0.29} (m_b) $\\&&&\\
   $E_l^{\Upsilon(4S)}>1.9$ GeV 
&\cite{Limosani:2005pi} 
&4.54
&$4.63\pm 0.43 ({\rm exp})  ^{+ 0.08} _{-0.12} ({\rm th}) _{ -0.31}  ^{+0.34} (m_b) $\\&&&\\
   $E_l^{\Upsilon(4S)}>2.0$ GeV 
&\cite{Aubert:2005mg}   
&4.17
&$4.24\pm 0.29 ({\rm exp})  ^{+ 0.11} _{-0.14} ({\rm th}) _{ -0.32}  ^{+0.37} (m_b) $\\&&&\\
  $E_l>2.0$ GeV; $S_h^{\max}<3.5$ GeV$^2$  
&\cite{Aubert:2005im}   
&4.18
&$4.24\pm 0.29 ({\rm exp})  ^{+ 0.12} _{-0.17} ({\rm th}) _{ -0.35}  ^{+0.40} (m_b) $\\&&&\\
   $E_l^{\Upsilon(4S)}>1.0$ GeV; $m_X<1.7$ GeV 
&\cite{Bizjak:2005hn}   
&4.21
&$4.31\pm 0.29 ({\rm exp})  ^{+ 0.06} _{-0.09} ({\rm th}) _{ -0.34}  ^{+0.40} (m_b) $\\&&&\\
  $E_l^{\Upsilon(4S)}>1.0$ GeV; $m_X<1.55$ GeV  
&\cite{Aubert:2007rb}   
&4.47
&$4.58\pm 0.22 ({\rm exp})  ^{+ 0.10} _{-0.12} ({\rm th}) _{ -0.44}  ^{+0.54} (m_b) $\\&&&\\
    \hline
  \end{tabular}
  \label{Vub_tab}
\end{table}
Note that the uncertainty in the total width is dominated by $m_b$ and it is therefore fully correlated with the parametric uncertainty associated with $m_b$ in $R_{\rm cut}$. This is taken into account in the parametric uncertainty quoted in Table~\ref{Vub_tab}. Because the effect of changing $m_b$ on $R_{\rm cut}$ acts in the same direction as in the total width, their product, which enters the determination of $|V_{ub}|$ in (\ref{DeltaB}), is highly sensitive to $m_b$. This is clearly reflected in the parametric uncertainty quoted in the table. It is also illustrated in figure \ref{fig:Vub_mass_dependence}.

Examining the values of $|V_{ub}|$ corresponding to different cuts one observes very good agreement. Even ignoring the theoretical and parametric uncertainties (which are correlated), they all agree very well. There is one case where the agreement is not as striking: this is the CLEO result of Ref.~\cite{Bornheim:2002du} where the central value falls below all other determinations; note however the large experimental error quoted.  

Averaging the results\footnote{Note that in this average we have neglected several correlations. A more precise updated average is being prepared by the HFAG.}
 in table \ref{Vub_tab} we obtain
\begin{equation}
\label{DGE_ave}
|V_{ub}|=\Big(4.30\pm 0.16 ({\rm exp}) ^{+0.09}_{-0.13}({\rm th}) _{- 0.34}^{ +0.39} (m_b) \Big)\cdot 10^{-3}\,.
\end{equation}
This result can be compared with other theoretical methods used to compute the partial widths (we only refer here to the two other methods that have been discussed above; HFAG \cite{HFAG_PDG08} presents additional results). 
The HQE--based parametrization of Ref.~\cite{Gambino:2007rp} yields an average~\cite{HFAG_PDG08}
\begin{equation}
\label{Gambino}
|V_{ub}|= \Big(3.94 \pm 0.15 ({\rm exp}) ^{+ 0.20}_{ - 0.23} ({\rm th})\Big)\cdot 10^{-3}\,,
\end{equation}
while the shape--function approach of Refs.~\cite{Bosch:2004th,Lange:2005yw} yields~\cite{Neubert:2008cp} 
 \begin{equation}
\label{BLNP}
|V_{ub}|= \Big(4.31 \pm 0.17 ({\rm exp}) \pm  0.35 ({\rm th})\Big)
\cdot 10^{-3}\,,
\end{equation}
where in all cases we quoted the numbers corresponding to the central value of Eq.~(\ref{m_b_PDG}) --- because of the strong $m_b$ dependence this requirement is essential for any valuable comparison.
For the average value one finds very good agreement between (\ref{DGE_ave}) and (\ref{BLNP}) and compatibility with (\ref{Gambino}). 
We note that for the $m_X$--based cuts there is better agreement between the methods of Ref.~\cite{Gambino:2007rp} and Ref.~\cite{Neubert:2008cp}, which both yield somewhat lower central values for $|V_{ub}|$ ($|V_{ub}|\simeq 4.0 \cdot 10^{-3}$ ~\cite{HFAG_PDG08,Neubert:2008cp} for the above $m_b$) as compared to the DGE approach.

\section{Conclusions}

I have given an overview of the main theoretical approaches used to compute the triple differential spectra in order to extract of $|V_{ub}|$ from data. I have mainly emphasized the conceptual differences and the relations between the approaches, but I also reported briefly on their status, their formal accuracy and their particular sources of uncertainty.

It is evident that despite making different approximations, the various determinations are consistent with each other. Add to that the remarkable consistency between different measurements that use different kinematics cuts --- which provides a valuable confirmation for the theoretical description of the spectrum --- the conclusion is clear: the inclusive determination of $|V_{ub}|$ is robust. This puts us on firm grounds coming to examine the consistency of the CKM mechanism.
 
Finally, the single most important source of uncertainty in the inclusive determination of $|V_{ub}|$ is the b-quark mass. The dependence on the mass is extremely high owing to the fact that both the total width and the cut--dependence increase with increasing $m_b$. The effect this has on
$|V_{ub}|$ is shown in figure \ref{fig:Vub_mass_dependence}.
Clearly, improving our knowledge of the b-quark mass would directly translate into more precise $|V_{ub}|$.

Before concluding I find it appropriate to add a few words about the field. Beyond their obvious significance to phenomenology, inclusive B decays are a remarkable source of interesting theoretical problems in QCD. We have only scratched the surface of this exciting field in this talk. 

Inclusive decay are also very challenging experimentally, and although the experimental issues have not been mentioned here, it is obvious that
there would not have been much point in giving this talk if not for the remarkable achievements of the B factories in this area. 
The on-going discussion between theory and experiment has also been extremely fruitful, and I would like to thank all those who have contributed to that.

\end{document}